\documentclass[fleqn,usenatbib]{mnras}

\usepackage{graphicx}
\usepackage{amssymb}
\usepackage{txfonts}
\usepackage{amsfonts}%
\usepackage{gensymb}
\newcommand{\modif}[1]{\color{black}#1\color{black}}

\title[2025 Draconid Forecast]{Model Predictions for the 2025 October Draconid Outburst}

\author[A. Egal et al.]{
Auriane Egal$^{1,2,3}$\thanks{E-mail: aegal@uwo.ca},
Paul Wiegert$^{2,4}$,
Danielle E. Moser$^{5}$,
Peter G. Brown$^{2,4}$,
Margaret Campbell-Brown$^{2,4}$\\
$^{1}$Plan\'etarium de Montr\'eal, Espace pour la Vie, 4801 av. Pierre-de Coubertin, Montr\'eal, Qu\'ebec H1V 3V4, Canada\\
$^{2}$Department of Physics and Astronomy, University of Western Ontario, London, ON N6A 3K7, Canada\\
$^{3}$Observatoire de Paris, PSL Research University, CNRS, Sorbonne Universit\'e, Universit\'e Lille, Paris, France\\
$^{4}$Institute for Earth and Space Exploration (IESX), University of Western Ontario, London, ON N6A 3K7, Canada\\
$^{5}$NASA Meteoroid Environment Office, Marshall Space Flight Center, Huntsville, AL 35812, USA
}

\date{Accepted XXX. Received YYY; in original form ZZZ}
\pubyear{2025}

\begin{document}

\maketitle

\begin{abstract}

The October Draconid meteor shower, produced by comet 21P/Giacobini–Zinner, is notorious for rare but intense outbursts, some exceeding rates of $\sim$10$^{4}$ meteors per hour. In 2025, Earth will encounter young trails ejected by the comet in 2005 and 2012, producing a meteor outburst and providing a rare opportunity to probe their structure and benchmark meteoroid stream models.
We present predictions from three independent dynamical models (NIMS, MSFC, Sisyphus), calibrated against updated activity profiles including the newly observed 2019 and 2024 outbursts. All simulations predict enhanced activity on 2025 October 8, dominated by faint meteors ($m<10^{-5}$ kg; $+$4 mag and fainter) primarily detectable by radar. Our best estimate is a radar outburst near 15:00–16:00 UT, driven mainly by the 2012 trail with a possible minor contribution from 2005.
The 2025 Draconids may represent one of the strongest radar dominated outbursts of the decade. Coordinated observing campaigns, especially radar measurements across the Northern Hemisphere and optical coverage from Asia, will be essential to validate these forecasts, constrain the dust environment of comet 21P, and improve future predictions of young meteoroid trails.
 
\end{abstract}


\section{Introduction}

The October Draconid (009 DRA) meteor shower is an episodic annual shower active around October 9. Most years it is weak, with Zenithal Hourly Rates (ZHR) below 1, but on occasion it produces spectacular outbursts and storms reaching hundreds to thousands of meteors per hour \citep[e.g.,][]{Watson1934,Lovell1947,Kresak1975,Ye2014}. The most dramatic storms in 1933 and 1946 reached ZHRs near 10$^{4}$ \citep{Jenniskens2006}, and early studies revealed the meteoroids to be among the weakest known \citep{Jacchia1950,Borovicka2007}.  

The Draconid stream shows clear evidence of mass segregation, with a mass index that varies strongly both between apparitions and with solar longitude \citep{Vida2020}. Some outbursts have been detected almost exclusively by radar (e.g., 1999 and 2012; \citealt{Campbell-Brown2021}), complicating comparisons between instruments and suggesting that past radar-dominated events may have gone unnoticed. A striking example is the 1999 outburst, only identified two decades later in archival radar data \citep{Egal2019}. The unpredictability and occasionally high fluxes of the Draconids, combined with the difficulty of their instrumental detection, make this shower a noteworthy hazard to spacecraft \citep{Moorhead2025}.  

\modif{The parent body of the Draconids, comet 21P/Giacobini–Zinner, is a Jupiter-family comet with a radius of 1–2 km \citep{Leibowitz1986,Singh1997,Lamy2004,Pittichova2008}. Its orbital evolution over the past two centuries has been unusually chaotic \citep{Marsden1971}, with multiple close encounters with Jupiter substantially altering the orbit, including one in 1898 that preceded its discovery \citep{Vaubaillon2011}. In addition, 21P has shown abrupt changes in nongravitational forces, including a marked shift between its 1959 and 1965 apparitions \citep{Yeomans1989,Sekanina1985,Ehlert2019}. }

The comet’s chaotic orbital history, combined with the observational challenges of the Draconid shower, makes accurate forecasting difficult. Even with large-scale dynamical simulations, predicting shower intensity remains uncertain, particularly for radar-dominated outbursts of small particles \citep{Egal2019}. The 2025 return offers a timely opportunity to test model performance against well-constrained recent apparitions. In this work, we present new forecasts based on three independent stream models, calibrated with updated activity profiles, including the newly measured 2019 and 2024 outbursts. Our modeling approach is described in Section~\ref{sec:models}, observational constraints in Section~\ref{sec:calibration}, and predictions for 2025 are provided in Section~\ref{sec:results}.

\section{Stream models}\label{sec:models}

\modif{We applied three independent meteoroid stream models to forecast the 2025 Draconid return: NIMS, MSFC, and Sisyphus. In all cases, the simulations follow dust released by comet 21P and identify Earth-approaching particles. Potential impactors were retained if they crossed the ecliptic plane within a distance $\Delta X$ and a time $\Delta T$ of Earth’s position, and were then used to generate simulated activity profiles for comparison with observations. Each model applied this common framework with distinct assumptions for the comet’s ephemerides and particle ejection, as described below.}

\begin{enumerate}

\item The Numerical Integration of Meteoroid Streams (NIMS; \citealt{Egal2019}) simulates dust released by comet 21P since the mid-19th century. A total of 14.42 million particles were ejected using the model of \citet{Crifo1997}, spanning sizes from $\sim$100~$\mu$m to 10~cm and assuming a density of 300~kg~m$^{-3}$ \citep{Borovicka2007}. Their subsequent motion was integrated under planetary perturbations and radiation forces. Potential impactors were retained if they crossed the ecliptic plane within $\Delta X=0.008$~au and $\Delta T=10$~days of Earth’s position. Particles were weighted according to the comet’s dust production profile measured in 2018 \citep{Egal2019}, following the procedure of \citet{Egal2020}. For the first time, separate modelled activity profiles were generated for radar-detectable meteors ($+$4 to $+$8 mag) and for optical meteors (brighter than $+$6 mag).

\item The NASA MSFC Meteoroid Stream Model (MSFC; \citealt{Moser2004, Moser2008}) simulates dust production by comet 21P since 1608. A total of 336 million particles were ejected following the model of \citet{Jones1996b}, with sizes between $\sim$100~$\mu$m and 10~cm and a density of 1000~kg~m$^{-3}$. Particles were integrated under gravitational and radiation forces, and those approaching Earth within $\Delta X=0.01$~au and $\Delta T=7$~days were selected for analysis.

\item The Sisyphus model (this work) focuses on trails released during the 2005 and 2012 perihelion passages. For each passage, 10,000 meteoroids of assumed density 300~kg~m$^{-3}$ were ejected using the velocity distribution of \citet{brojon98}. Their motion was integrated under planetary perturbations and radiation forces. Meteoroids were considered part of the 2025 Draconid shower if they passed Earth with $\Delta T= 7$~days and $\Delta X= 0.02$~au.

\end{enumerate}

The chaotic orbital history of comet 21P, including a close encounter with Jupiter in 1898 and limited pre-1966 astrometry \citep{Vaubaillon2011, Yeomans1989}, prevents direct long-term integration from present-day elements. \modif{To address these uncertainties, the ephemeris of each apparition was computed by integrating the comet's motion forward or backward from the nearest accurate orbital solution available for that epoch, rather than from the most recent orbit. We used JPL Horizons\footnote{\url{https://ssd.jpl.nasa.gov/horizons}, accessed June 2024} orbital solutions spanning 1900–2017 (SAO series for 1900–1998; K054/18 for 2006; K123/6 for 2013; K253/3 for 2017). } For the Sisyphus model, we adopted the DE440 ephemerides, with initial conditions and non-gravitational parameters for 21P obtained from JPL CNEOS at JDE 2458064.5 (7 Nov 2017).

\modif{Despite the large number of simulated particles, only a small fraction approach Earth’s orbit within the critical distance $\Delta X$ near the time of the shower \citep[e.g.,][]{Egal2019}. To address this, the $\Delta T$ parameter is introduced as an along-track uncertainty, effectively broadening each particle into a swarm of clones spread across $\pm\Delta T$ at its node and thereby increasing the number of retained impactors.

Restrictive selection criteria are necessary to capture the perturbed structure of the stream, but they inevitably reduce the statistical robustness of the forecasts. In this work, the $\Delta X,\Delta T$ values adopted for each model were chosen to reproduce the timing and intensity of well-constrained past Draconid outbursts (e.g. Section \ref{sec:calibration} for NIMS), and are consistent with thresholds successfully applied to streams from both Halley-type and Jupiter-family comets \citep[e.g.][]{Egal2020, egal_modeling_2023}.}

\section{Calibration with past Draconid activity}\label{sec:calibration}

\subsection{Observations}\label{sec:observations}

The observational characteristics of Draconid outbursts between 1926 and 2018 were summarized by \citet{Egal2019}. Since its discovery, the shower has produced four historic storms (1933, 1946, 1999, 2012) and several major outbursts (1985, 1998, 2005, 2011, 2018, 2019), separated by years of negligible activity.

Systematic radar measurements became available only after 1998 \citep{Campbell-Brown2021}. Since 2002, the CMOR radar has provided a nearly continuous and internally consistent dataset of Draconid activity. Revised flux computations by \citet{Campbell-Brown2021} substantially altered the reported intensities of several outbursts, underscoring the large uncertainties in radar flux determination \citep[potentially up to a factor of twenty for the Draconids, cf. ][]{Moorhead2024}. Flux calibration is further complicated by the shower’s highly variable mass index, which changes not only between apparitions but also during individual events \citep{Vida2020, Campbell-Brown2021}.

Figure \ref{fig:NIMS_postdictions} presents activity profiles from multi-instrument observations of all Draconid apparitions with significant flux enhancements since 1933. While radar data provide the only record of several outbursts (e.g., 1999, 2012, 2019), others (1998, 2011, 2018) were simultaneously observed with visual, optical, and radar techniques. Direct comparison reveals systematic offsets in peak timing (up to 30–45 minutes between optical systems) and discrepancies between optical and radar maxima, underscoring the importance of coordinated multi-instrument campaigns.

\modif{Two additional outbursts expand the record since 2018. In 2019, enhanced radio activity was detected by the Mohe radar and by CMOR, which measured an equivalent ZHR of $\sim$625 ($s=1.8$) at SL 194.734$\degree$ \citep{Li2022}. The smaller sub-peak near SL 194.56 detected in CMOR data is most likely caused by noise and is not present in the independent measurements of \citet{Li2022}. The main peak, detected by both systems, is in excellent agreement with predictions by \citet{Egal2019}. }On 2024 October 8, a weak unanticipated outburst was detected. GMN reported a peak ZHR of $\sim$16 hr$^{-1}$ at SL $195.08\pm0.05\degree$ \citep{CBET2024}, while Japanese radio and CMOR data indicated maxima about 45 minutes earlier and an hour later, respectively \citep{Rendtel2024}.

\subsection{Calibration}

\begin{figure*}
    \centering
    \includegraphics[width=0.33\linewidth]{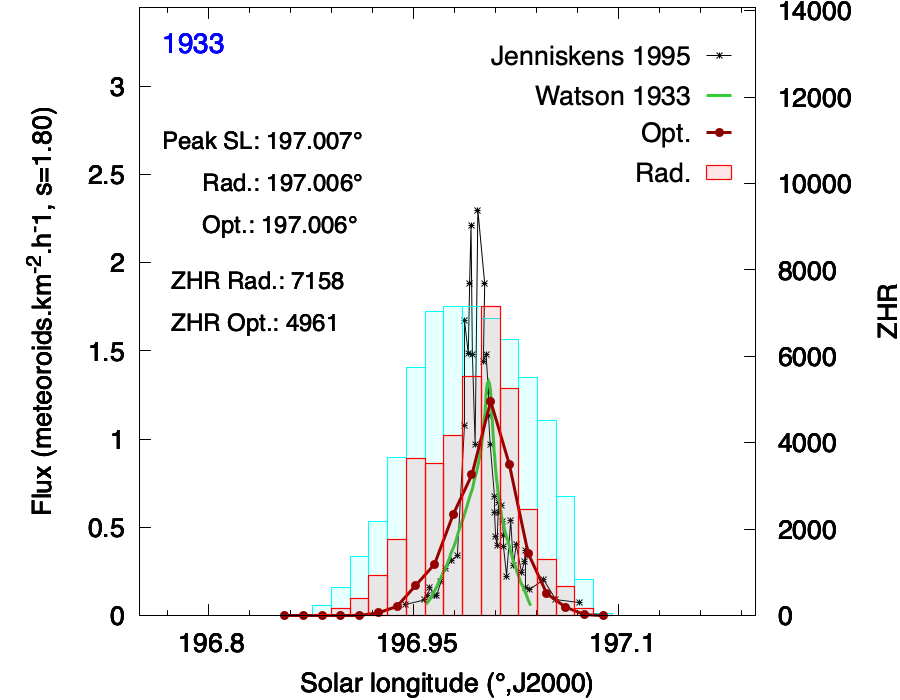}
    \includegraphics[width=0.33\linewidth]{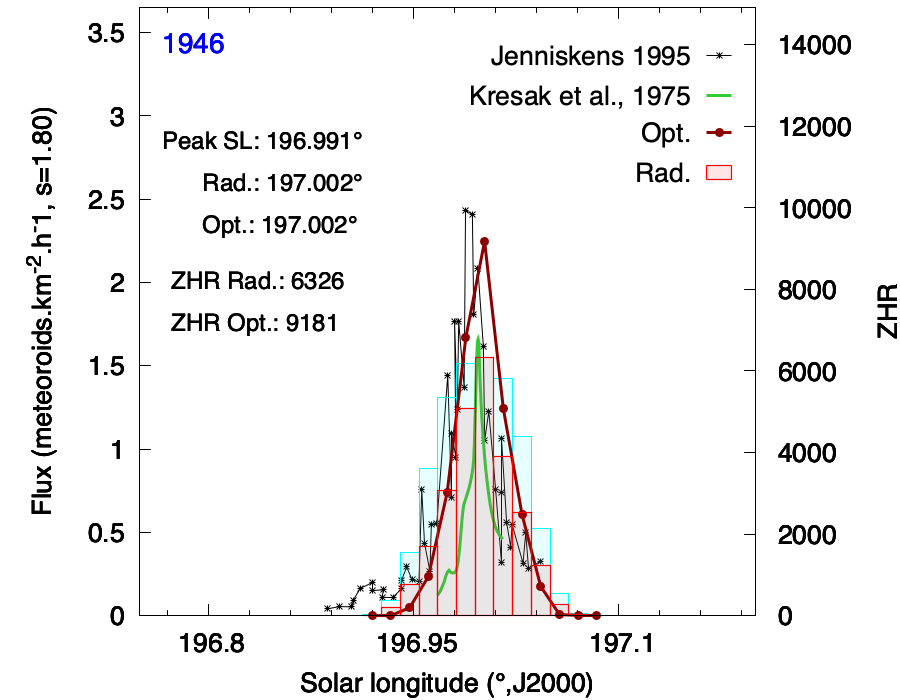}
    \includegraphics[width=0.33\linewidth]{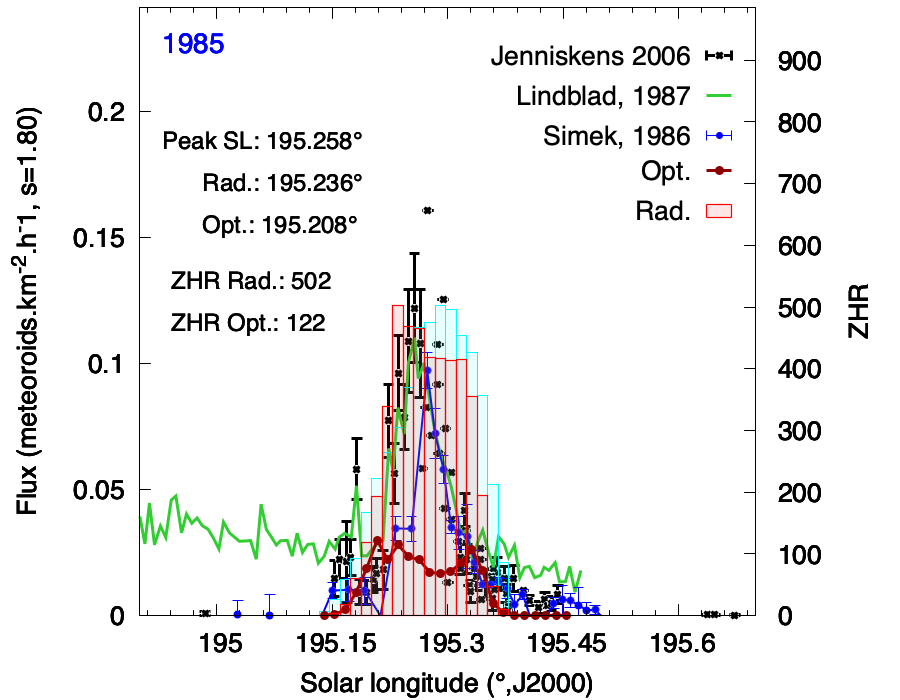}\\
    \includegraphics[width=0.33\linewidth]{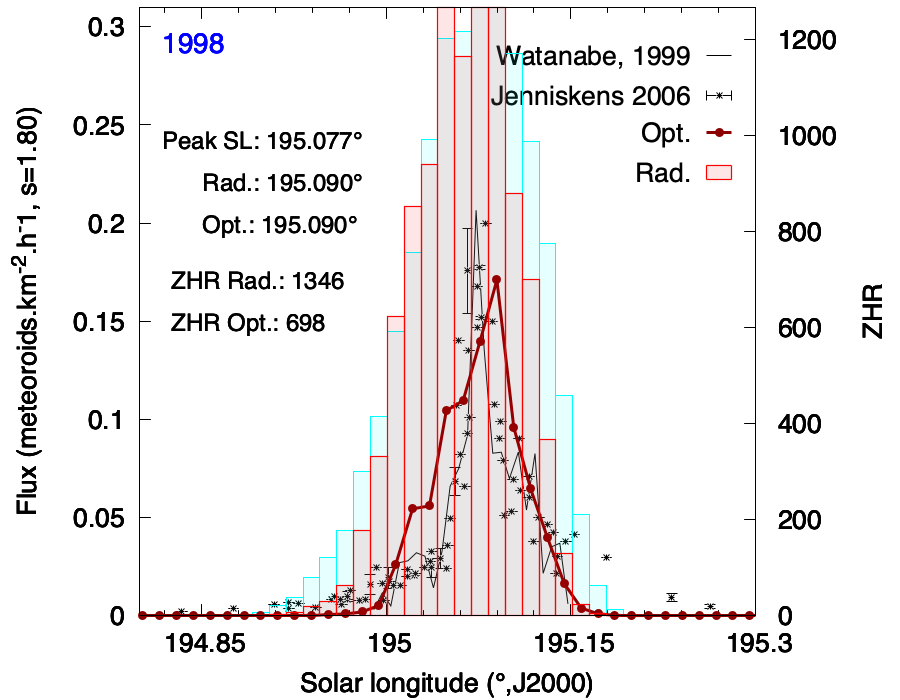}
    \includegraphics[width=0.33\linewidth]{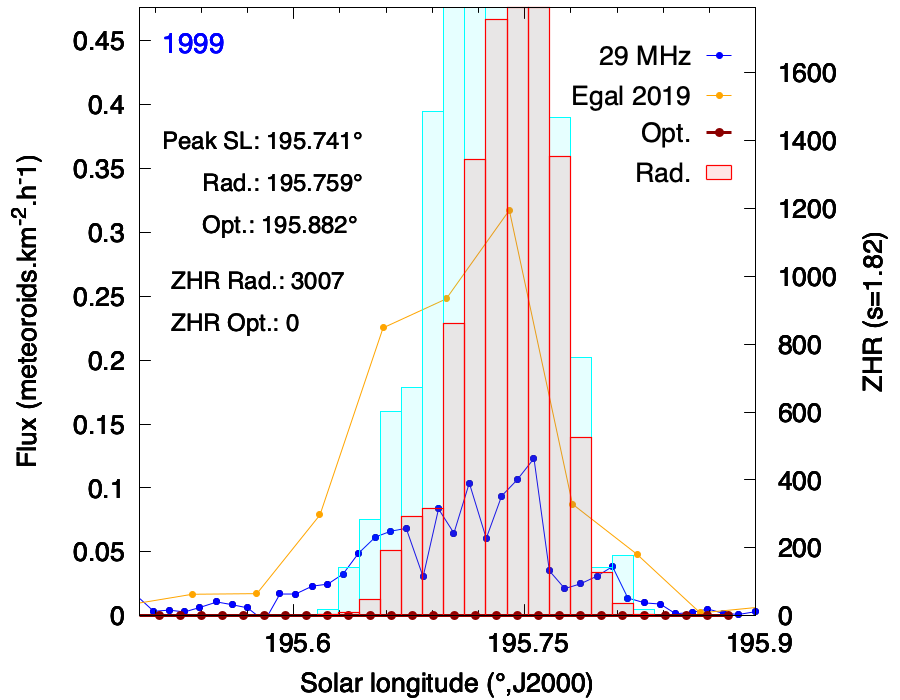}
    \includegraphics[width=0.33\linewidth]{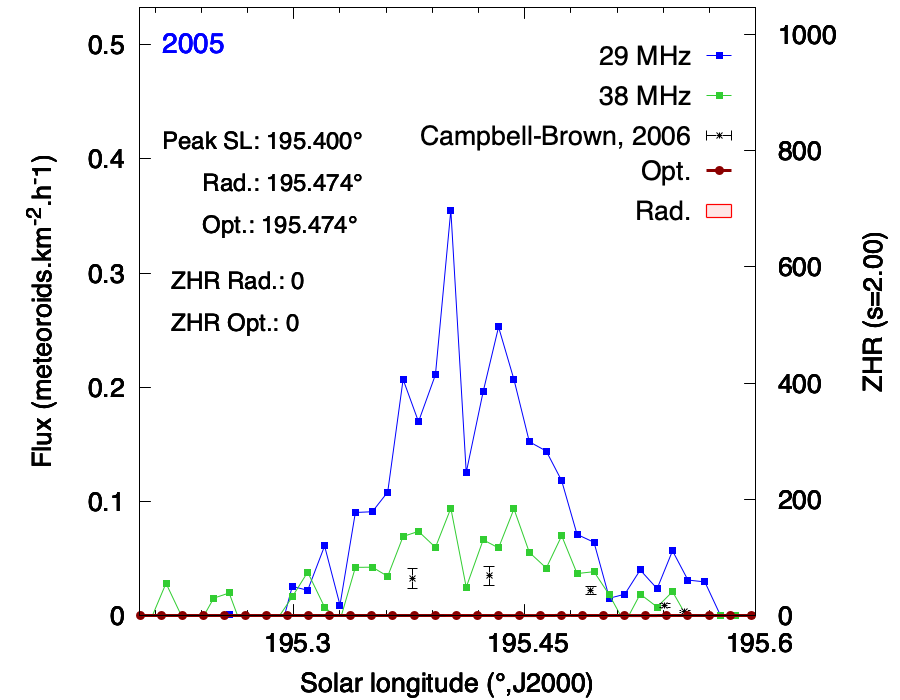}\\
    \includegraphics[width=0.33\linewidth]{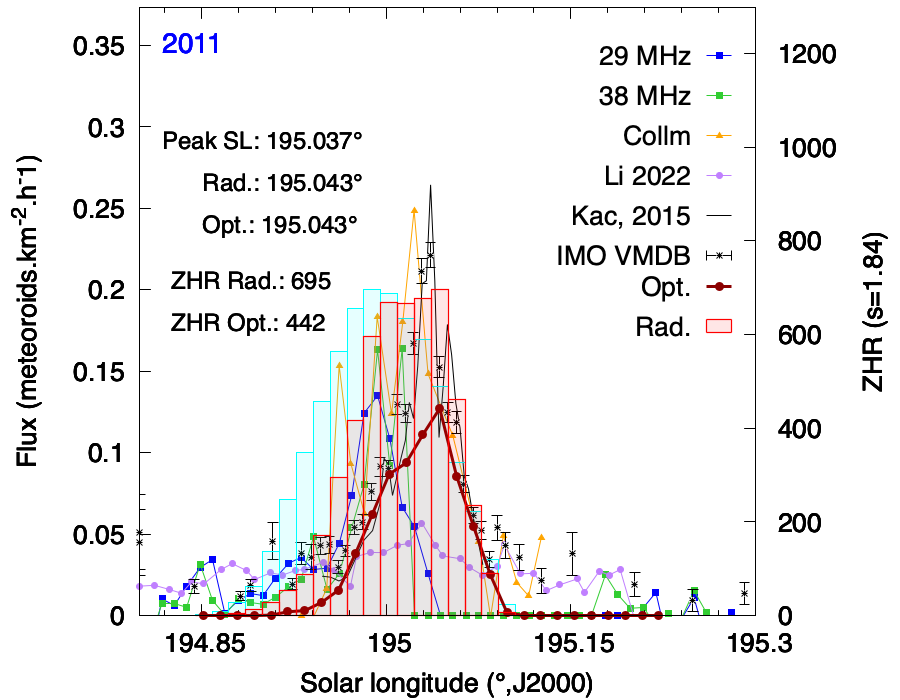}
    \includegraphics[width=0.33\linewidth]{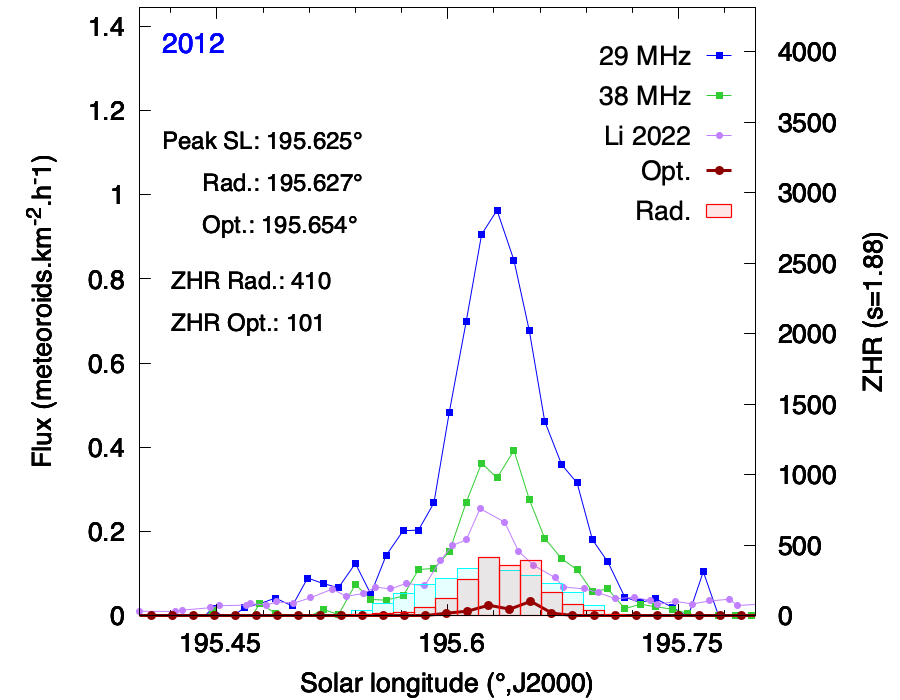}
    \includegraphics[width=0.33\linewidth]{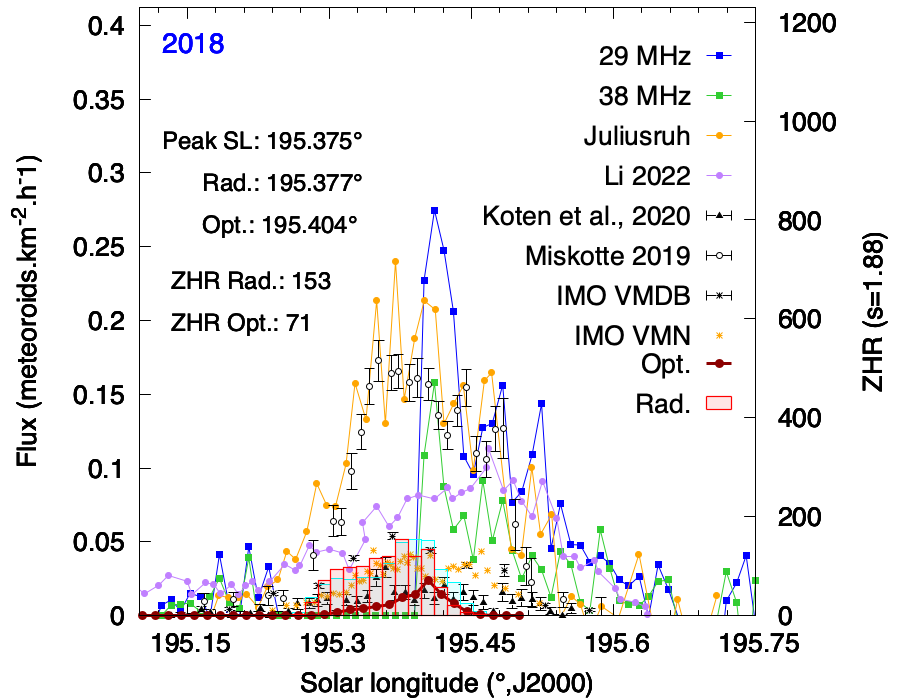}\\
    \includegraphics[width=0.33\linewidth]{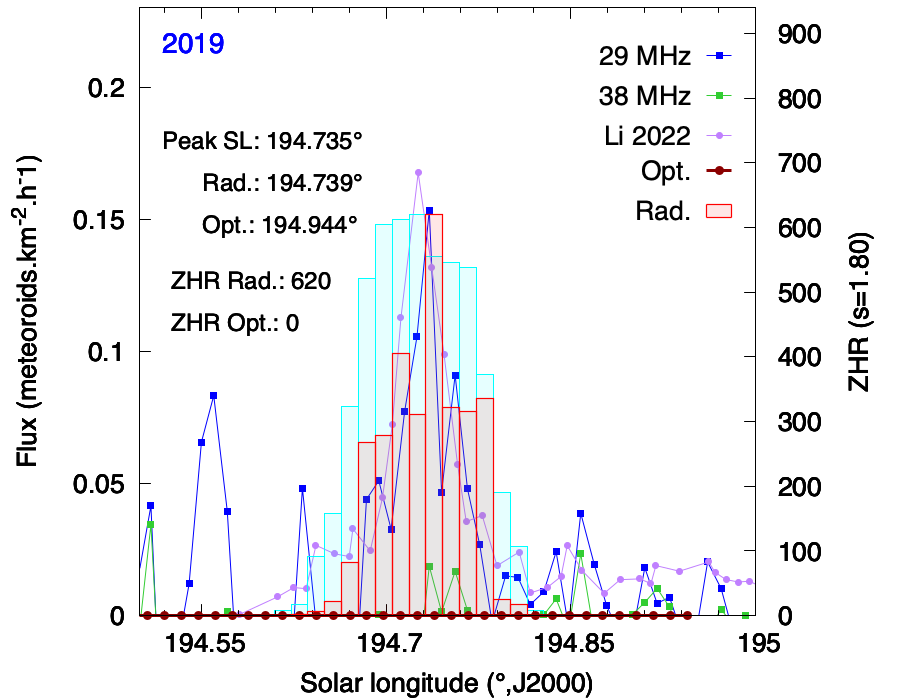}
    \includegraphics[width=0.33\linewidth]{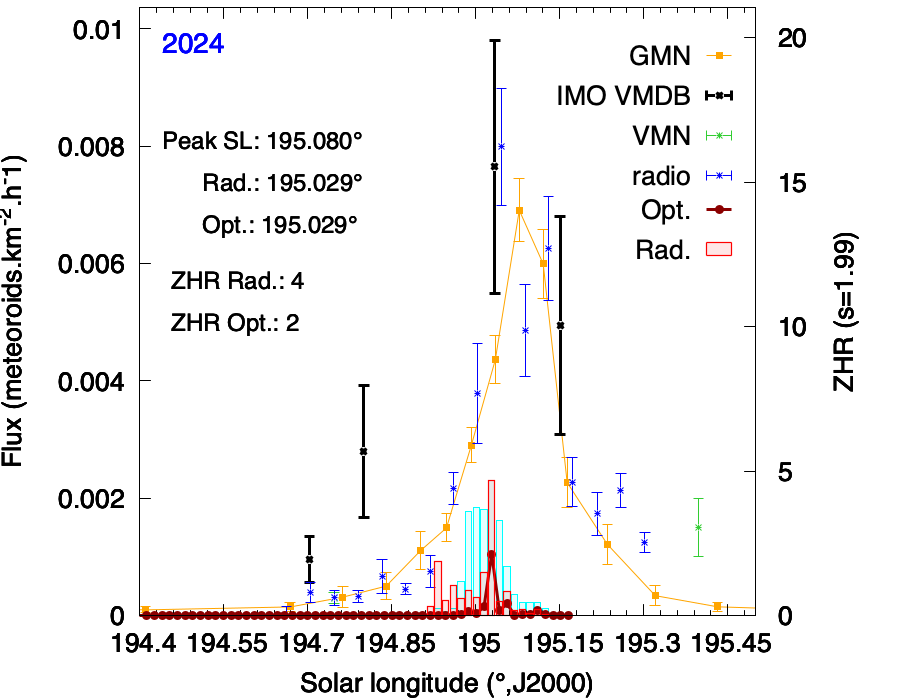}
    
    \caption{Postdiction of the main Draconid outbursts using NIMS. The weighted simulated activity profile for meteoroids producing meteors detectable at modern optical instrument sensitivities (solid red curve) and in the radar range (red boxes) are compared with visual, optical, and radio observations for each apparition. Observations reported in the literature are shown with their associated reference. Flux profile measurements by CMOR using either the 29 or 38 MHz radar system are also shown where such data are available.  ZHR to flux conversion used the mass index $s$ shown in each panel, or the default value $s=1.80$ from \citep{Campbell-Brown2021}. The raw simulated profile, obtained with minimal weighting and without calibration of the particle size distribution at ejection, is shown in cyan. For each plot, the observed peak solar longitude (Peak SL), along with the simulated time of maximum at radar sizes (Rad) and optical sizes (Opt) and the corresponding ZHR, are indicated. The flux is given to an equivalent Draconid peak magnitude of +6.5 which corresponds to a particle mass of 10$^{-6}$ kg.}
    \label{fig:NIMS_postdictions}
\end{figure*}

Our NIMS simulations were calibrated against the major outbursts shown in Figure \ref{fig:NIMS_postdictions}. For most apparitions, peak timing was reproduced within $\sim$30 minutes, and modeled fluxes were consistent with observations for six returns (1933, 1946, 1985, 1998, 2011, 2019). The weak 2024 enhancement was also reproduced, predicted about one hour earlier than the GMN optical maximum \citep{CBET2024} but in good agreement with radio measurements \citep{Rendtel2024}.  

Three returns (1999, 2012, 2018) proved more difficult to reproduce, though the results remain within observational uncertainties. Modeled fluxes diverged by factors of 2–8 depending on the dataset, reflecting substantial measurement challenges: the 1999 rates relied on a precursor radar with uncertain calibration \citep{Campbell-Brown2021}; the 2012 storm was produced by a trail ejected shortly after a major orbital alteration of 21P; and the 2018 outburst exhibited complex filamentary structure and rapid mass index variations ($s$ shifting from 1.74 to 2.32 within minutes; \citealt{Vida2020}). The 2005 return was missed entirely, as particles from the 1946–1953 trails were displaced beyond Earth’s orbit and excluded by our tight $\Delta X$ criterion.

Attempts to tune the model to these four apparitions degraded the fits for other years, underscoring the challenge of achieving a uniform calibration across the shower’s history. Given the uncertainty in radar flux measurements \citep{Moorhead2024}, we prioritized reproducing peak timing and intensity of the well-constrained outbursts over over calibrating to the uncertain cases. With this approach, the NIMS model reproduces the timing and structure of most Draconid outbursts and provides a consistent baseline for forecasting 2025.

\begin{figure*}
 \centering
    \includegraphics[width=0.97\linewidth]{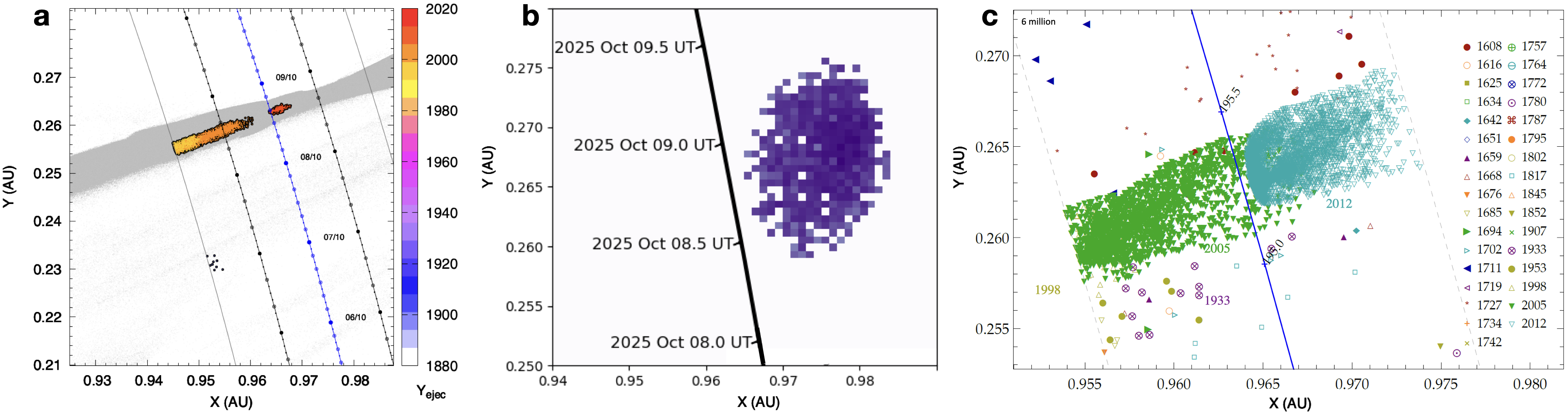}
    \includegraphics[height=4.36cm]{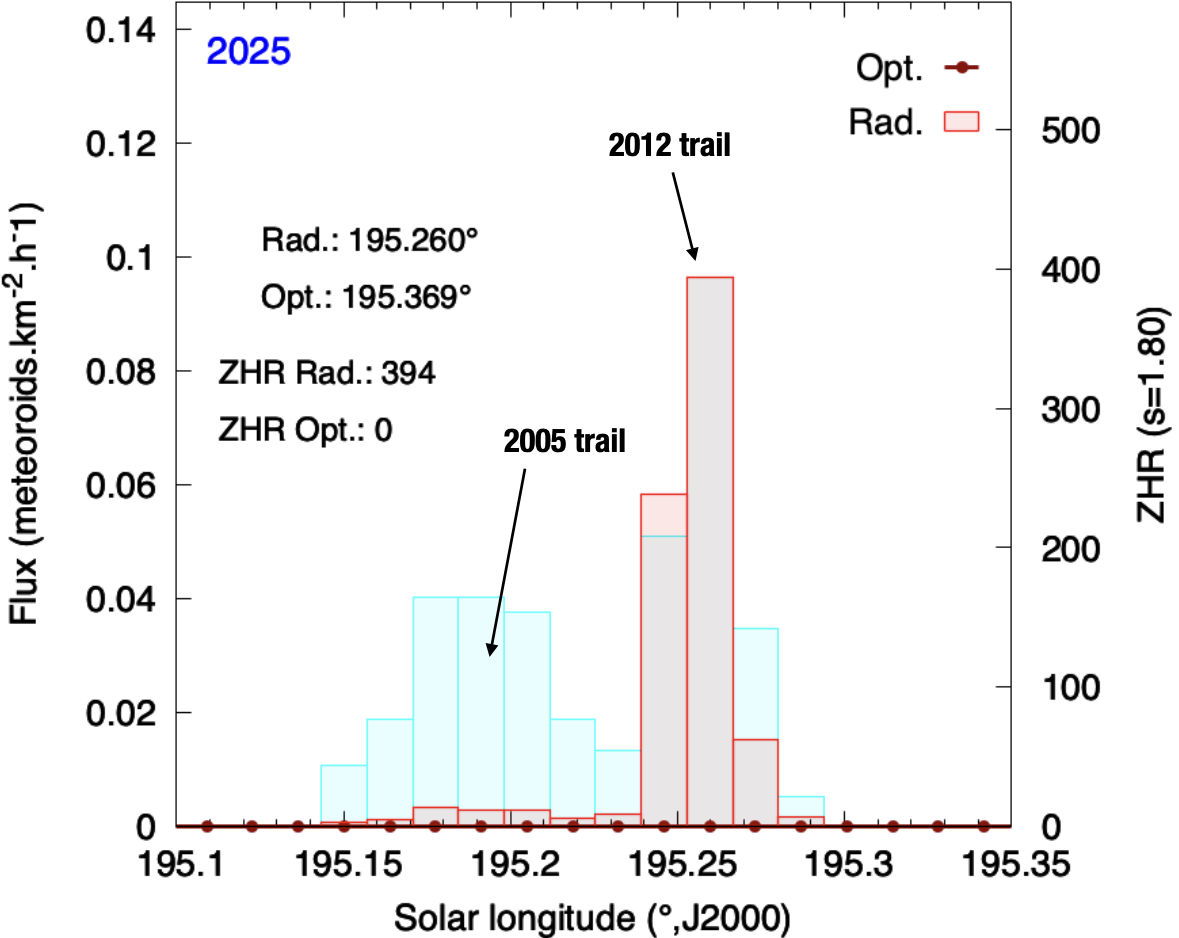}
    \includegraphics[height=4.40cm]{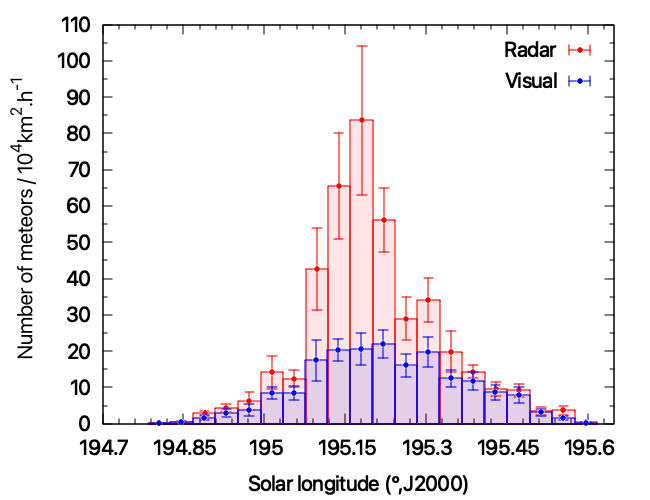}
    \includegraphics[height=4.40cm]{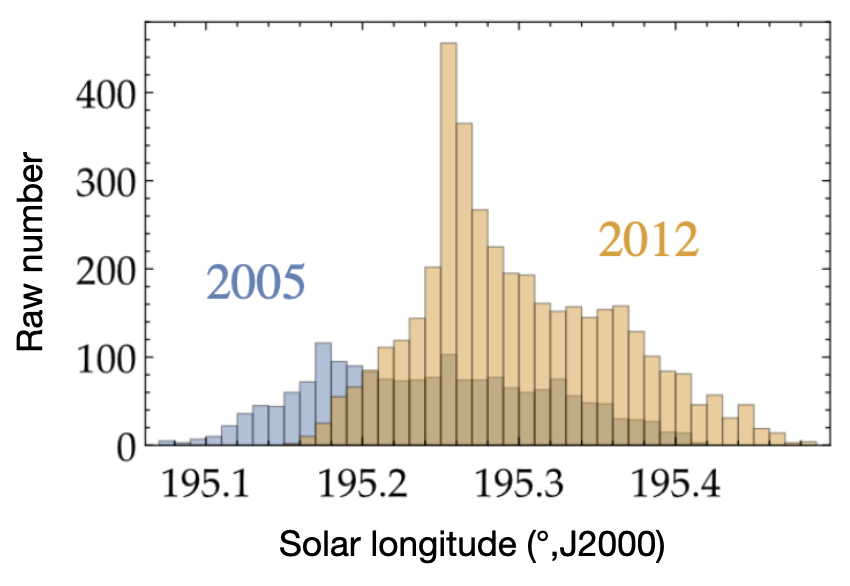}
   \caption{Simulated nodal-crossing locations of Draconid particles (top) and activity profile of the predicted 2025 outburst (bottom), using the NIMS (a), Sisyphus (b), and MSFC (c) models. Only particles crossing the ecliptic plane within the selection criteria ($\Delta X$, $\Delta T$) from Earth's position are shown in color. Particles released by comet 21P in 2005 and 2012 are shown separately in NIMS and MSFC simulated profiles.}
 \label{fig:2025_prediction}
\end{figure*}

\begin{table*}
 \centering
  \begin{tabular}{llllll}
  \hline
  Modeller & Trail &  SL ($\degree$) & Time (Oct. 8) & ZHR & Source \\
  \hline
  \hline
  Maslov & 2012 & 195.238 & 15:07 & 100-150 & \cite{Rendtel2025} \\
         & 1907-1953 & 194.663-195.053 & 01:07-10:36 & $<$50-60  & \cite{Maslov2025} \\
  \hline
  Ye & 1933 & 195.0 & 09:19 & $<$2018 rates & \cite{Ye2014}\\
  \hline
  Vaubaillon & 2012 & 195.269 & 15:52 & $<$50 & \cite{Rendtel2025} \\
             & 2005 & 195.272 & 15:56 & negligible & \cite{Rendtel2025} \\
  \hline
  Sato & 2012 & 195.257 & 15:34 & - & \cite{Rendtel2025} \\
  \hline
  Egal & 2012 & 195.252 & 15:27 & 950* & \cite{Egal2019}\\ 
       & 2005 & 195.178 & 13:39 & 950* & \cite{Egal2019}\\
       & 2012 & 195.256 & 15:32 & 400* & This work, NIMS \\
       & 2005 & 195.185 & 13:49 & 400* & This work, NIMS \\
   \hline
   Moser & 2012 & 195.255 & 15:31 & - & This work, MSFC \\
         & 2005 & 195.175 & 13:34 & - & This work, MSFC \\
         & 2005 \& 2012 fit & 195.264 & 15:45 & - & This work, MSFC \\
  \hline
   Wiegert & 2005 & 195.18/195.22& 13:42/14:40 & 45/12 & (radar/visual), This work, Sisyphus \\
         & 2012 & 195.18/195.14 & 13:42/12:43 & 35/13 & (radar/visual), This work, Sisyphus \\
  \hline
 \end{tabular}
 \caption{\label{table:forecast} 2025 Draconid forecasts performed by independent modelers. All the predicted activity is expected to occur on October 8, 2025. *: ZHR estimate based on the combined activity from the 2005 \& 2012 trails. }
\end{table*}

\section{2025 Forecast}\label{sec:results}

Figure~\ref{fig:2025_prediction} shows the predicted 2025 activity from the NIMS, MSFC, and Sisyphus models. We find that the trails ejected by comet 21P in 2005 and 2012 approach Earth’s orbit on October 8-9, together with more dispersed particles from earlier apparitions. Even with similar integration parameters, the models yield noticeably different trail structures and elongations. Depending on the adopted ejection and selection criteria, the 2005 trail either intersects Earth (MSFC and \citealt{Egal2019}), producing minor activity, or passes slightly beyond Earth’s orbit (NIMS and Sisyphus). The denser 2012 trail is offset from Earth in most models, but lies close enough that activity cannot be ruled out. Indeed, the 1966 trail was similarly offset in 2012, yet produced a radio storm of $\sim$3000 meteors h$^{-1}$ \citep{Ye2014,Campbell-Brown2021}. The predicted activity is therefore highly sensitive to the adopted ejection model and the ($\Delta X,\Delta T$) selection parameters. Our best estimate, calibrated against as many past outbursts as possible, is shown in Figure \ref{fig:2025_prediction}.  

All three models predict enhanced activity on October 8. NIMS yields a radar outburst of $\sim$400 meteors h$^{-1}$ at $\lambda_\odot=195.26^\circ$ (15:20 UT), dominated by the 2012 trail with an earlier, smaller contribution from 2005. MSFC results agree in timing and relative strength, also indicating dominance by sub-mm particles (masses $<10^{-6}$~kg) and hence radar-dominated activity. By contrast, the Sisyphus model favors a slightly earlier peak at $\lambda_\odot=195.18$–195.23$^\circ$, produced by mm–cm particles from the 2012 and 2005 trails. Interpreting Sisyphus rates (meteors per 10\,000~km$^{2}$ per hour) as a ZHR proxy gives modest radar activity (ZHR$_{\mathrm{eq}}\sim$90) and lower visual rates near 25 h$^{-1}$.

A comparison with other forecasts is given in Table \ref{table:forecast}. Most simulations predict enhanced radio activity from the 2012 trail on 2025 October 8, between 15:00–16:00 UT, with a possible 2005 contribution depending on meteoroid selection. The \citet{Egal2019} and NIMS predictions give the highest expected rates, while others suggest lower activity. The contribution of individual trails remains highly sensitive to the adopted ($\Delta X,\Delta T$), emphasizing the importance of 2025 observations for constraining future forecasts. The peak is expected during daytime in Europe and North America, making optical observations most favorable from Asia, while radio measurements across the Northern Hemisphere will be critical. 

\modif{Since the activity is expected to be dominated by faint meteors, enhanced video and telescopic techniques will be particularly valuable. Such systems can extend sensitivity beyond the naked-eye limit and help bridge the gap between optical and radar observations, providing more robust constraints on the shower’s mass distribution. } Unfortunately, optical observations may be hindered by the full Moon occurring the day prior to the peak, on October 7. 

\section{Conclusions}

Our simulations indicate that the October 8, 2025, Draconid return is likely to produce a moderate to strong radio outburst, primarily originating from the dust trail ejected by comet 21P/Giacobini-Zinner in 2012. A minor contribution from the 2005 trail and/or mm–cm sized particles is also possible. While all models predict enhanced activity on October 8, the exact timing and relative strength of the peaks differ significantly, underlining the sensitivity of the forecasts to the adopted ejection models and particle selection criteria.

These differences highlight two priorities for improving future forecasts: (1) refining the ephemeris of comet 21P to reduce uncertainties in the nodal-crossing location of the stream, and (2) obtaining more reliable Draconid flux measurements through coordinated multi-instrument observing campaigns, with particular emphasis on accurately determining the shower's mass index.

The 2025 return offers a particularly valuable opportunity to test and validate both the ejection model and the particle selection parameters. The predicted activity originates from young trails with well-constrained cometary apparitions, enabling a more direct comparison between model predictions and observations. Although the peak occurs during daytime in Europe and North America, it will be observable under night-time conditions in Asia and detectable with radio instruments across the Northern Hemisphere. High-quality measurements in 2025 will thus be crucial both for advancing our understanding of the Draconid stream and for benchmarking future forecasts.

\section*{Acknowledgements}
AEg, PBr, PWi, and MCa were supported by the NASA Meteoroid Environment Office under cooperative agreement 80NSSC24M0060.

\section*{Data Availability Statement}

The data underlying this article will be shared on reasonable request to the corresponding author.

\bibliographystyle{mnras}
\bibliography{references}

\begin{thebibliography}{}
\makeatletter
\relax
\def\mn@urlcharsother{\let\do\@makeother \do\$\do\&\do\#\do\^\do\_\do\%\do\~}
\def\mn@doi{\begingroup\mn@urlcharsother \@ifnextchar [ {\mn@doi@}
  {\mn@doi@[]}}
\def\mn@doi@[#1]#2{\def\@tempa{#1}\ifx\@tempa\@empty \href
  {http://dx.doi.org/#2} {doi:#2}\else \href {http://dx.doi.org/#2} {#1}\fi
  \endgroup}
\def\mn@eprint#1#2{\mn@eprint@#1:#2::\@nil}
\def\mn@eprint@arXiv#1{\href {http://arxiv.org/abs/#1} {{\tt arXiv:#1}}}
\def\mn@eprint@dblp#1{\href {http://dblp.uni-trier.de/rec/bibtex/#1.xml}
  {dblp:#1}}
\def\mn@eprint@#1:#2:#3:#4\@nil{\def\@tempa {#1}\def\@tempb {#2}\def\@tempc
  {#3}\ifx \@tempc \@empty \let \@tempc \@tempb \let \@tempb \@tempa \fi \ifx
  \@tempb \@empty \def\@tempb {arXiv}\fi \@ifundefined
  {mn@eprint@\@tempb}{\@tempb:\@tempc}{\expandafter \expandafter \csname
  mn@eprint@\@tempb\endcsname \expandafter{\@tempc}}}

\bibitem[\protect\citeauthoryear{{Borovi{\v c}ka}, {Spurn{\'y}}  \&
  {Koten}}{{Borovi{\v c}ka} et~al.}{2007}]{Borovicka2007}
{Borovi{\v c}ka} J.,  {Spurn{\'y}} P.,   {Koten} P.,  2007, \mn@doi [Astronomy
  and Astrophysics] {10.1051/0004-6361:20078131}, \href
  {http://cdsads.u-strasbg.fr/abs/2007A%26A...473..661B} {473, 661}

\bibitem[\protect\citeauthoryear{{Brown} \& {Jones}}{{Brown} \&
  {Jones}}{1998}]{brojon98}
{Brown} P.,  {Jones} J.,  1998, \mn@doi [Icarus] {10.1006/icar.1998.5920},
  \href {http://adsabs.harvard.edu/abs/1998Icar..133...36B} {133, 36}

\bibitem[\protect\citeauthoryear{{Campbell-Brown}, {Stober}, {Jacobi}, {Kero},
  {Kozlovsky}  \& {Lester}}{{Campbell-Brown} et~al.}{2021}]{Campbell-Brown2021}
{Campbell-Brown} M.~D.,  {Stober} G.,  {Jacobi} C.,  {Kero} J.,  {Kozlovsky}
  A.,   {Lester} M.,  2021, \mn@doi [\mnras] {10.1093/mnras/stab2174}, \href
  {https://ui.adsabs.harvard.edu/abs/2021MNRAS.507..852C} {507, 852}

\bibitem[\protect\citeauthoryear{{Crifo} \& {Rodionov}}{{Crifo} \&
  {Rodionov}}{1997}]{Crifo1997}
{Crifo} J.~F.,  {Rodionov} A.~V.,  1997, \mn@doi [Icarus]
  {10.1006/icar.1997.5690}, \href
  {http://cdsads.u-strasbg.fr/abs/1997Icar..127..319C} {127, 319}

\bibitem[\protect\citeauthoryear{{Egal}, {Wiegert}, {Brown}, {Moser},
  {Campbell-Brown}, {Moorhead}, {Ehlert}  \& {Moticska}}{{Egal}
  et~al.}{2019}]{Egal2019}
{Egal} A.,  {Wiegert} P.,  {Brown} P.~G.,  {Moser} D.~E.,  {Campbell-Brown} M.,
   {Moorhead} A.,  {Ehlert} S.,   {Moticska} N.,  2019, \mn@doi [\icarus]
  {10.1016/j.icarus.2019.04.021}, \href
  {https://ui.adsabs.harvard.edu/abs/2019Icar..330..123E} {330, 123}

\bibitem[\protect\citeauthoryear{{Egal}, {Wiegert}, {Brown}, {Campbell-Brown}
  \& {Vida}}{{Egal} et~al.}{2020}]{Egal2020}
{Egal} A.,  {Wiegert} P.,  {Brown} P.~G.,  {Campbell-Brown} M.,   {Vida} D.,
  2020, \mn@doi [\aap] {10.1051/0004-6361/202038953}, \href
  {https://ui.adsabs.harvard.edu/abs/2020A&A...642A.120E} {642, A120}

\bibitem[\protect\citeauthoryear{Egal, Wiegert, Brown  \& Vida}{Egal
  et~al.}{2023}]{egal_modeling_2023}
Egal A.,  Wiegert P.~A.,  Brown P.~G.,   Vida D.,  2023, \mn@doi [The
  Astrophysical Journal] {10.3847/1538-4357/acb93a}, 949, 96

\bibitem[\protect\citeauthoryear{{Ehlert}, {Moticska}  \& {Egal}}{{Ehlert}
  et~al.}{2019}]{Ehlert2019}
{Ehlert} S.,  {Moticska} N.,   {Egal} A.,  2019, \mn@doi [\aj]
  {10.3847/1538-3881/ab1d50}, \href
  {https://ui.adsabs.harvard.edu/abs/2019AJ....158....7E} {158, 7}

\bibitem[\protect\citeauthoryear{Jacchia, Kopal  \& Millman}{Jacchia
  et~al.}{1950}]{Jacchia1950}
Jacchia L.,  Kopal Z.,   Millman P.,  1950, The Astrophysical Journal, 111, 104

\bibitem[\protect\citeauthoryear{{Jenniskens}}{{Jenniskens}}{2006}]{Jenniskens2006}
{Jenniskens} P.,  2006, {Meteor Showers and their Parent Comets}

\bibitem[\protect\citeauthoryear{{Jones} \& {Brown}}{{Jones} \&
  {Brown}}{1996}]{Jones1996b}
{Jones} J.,  {Brown} P.,  1996, in {Gustafson} B. A.~S.,  {Hanner} M.~S.,  eds,
   Astronomical Society of the Pacific Conference Series Vol. 104, IAU
  Colloquium 150: Physics, Chemistry, and Dynamics of Interplanetary Dust.
  p.~137

\bibitem[\protect\citeauthoryear{{Kresak} \& {Slancikova}}{{Kresak} \&
  {Slancikova}}{1975}]{Kresak1975}
{Kresak} L.,  {Slancikova} J.,  1975, Bulletin of the Astronomical Institutes
  of Czechoslovakia, \href {http://adsabs.harvard.edu/abs/1975BAICz..26..327K}
  {26, 327}

\bibitem[\protect\citeauthoryear{{Lamy}, {Toth}, {Fernandez}  \&
  {Weaver}}{{Lamy} et~al.}{2004}]{Lamy2004}
{Lamy} P.~L.,  {Toth} I.,  {Fernandez} Y.~R.,   {Weaver} H.~A.,  2004, {The
  sizes, shapes, albedos, and colors of cometary nuclei}.
pp 223--264

\bibitem[\protect\citeauthoryear{{Leibowitz} \& {Brosch}}{{Leibowitz} \&
  {Brosch}}{1986}]{Leibowitz1986}
{Leibowitz} E.~M.,  {Brosch} N.,  1986, \mn@doi [Icarus]
  {10.1016/0019-1035(86)90049-7}, \href
  {http://cdsads.u-strasbg.fr/abs/1986Icar...68..430L} {68, 430}

\bibitem[\protect\citeauthoryear{{Li}, {Li}, {Hu}, {Zhao}, {Sun}, {Xie}, {Yang}
   \& {Ning}}{{Li} et~al.}{2022}]{Li2022}
{Li} Y.,  {Li} G.,  {Hu} L.,  {Zhao} X.,  {Sun} W.,  {Xie} H.,  {Yang} S.,
  {Ning} B.,  2022, \mn@doi [\mnras] {10.1093/mnras/stac2589}, \href
  {https://ui.adsabs.harvard.edu/abs/2022MNRAS.516.5538L} {516, 5538}

\bibitem[\protect\citeauthoryear{{Lovell}, {Banwell}  \& {Clegg}}{{Lovell}
  et~al.}{1947}]{Lovell1947}
{Lovell} A.~C.~B.,  {Banwell} C.~J.,   {Clegg} J.~A.,  1947, \mn@doi [Monthly
  Notices of the RAS] {10.1093/mnras/107.2.164}, \href
  {http://adsabs.harvard.edu/abs/1947MNRAS.107..164L} {107, 164}

\bibitem[\protect\citeauthoryear{{Marsden} \& {Sekanina}}{{Marsden} \&
  {Sekanina}}{1971}]{Marsden1971}
{Marsden} B.~G.,  {Sekanina} Z.,  1971, \mn@doi [Astronomical Journal]
  {10.1086/111232}, \href {http://cdsads.u-strasbg.fr/abs/1971AJ.....76.1135M}
  {76, 1135}

\bibitem[\protect\citeauthoryear{Maslov}{Maslov}{2025}]{Maslov2025}
Maslov M.,  2025, Draconids 1901--2100: activity predictions, \url
  {http://feraj.ru/Radiants/Predictions/1901-2100eng/Draconids1901-2100predeng.html}

\bibitem[\protect\citeauthoryear{Moorhead, Campbell-Brown  \& Brown}{Moorhead
  et~al.}{2024}]{Moorhead2024}
Moorhead A.~V.,  Campbell-Brown M.~D.,   Brown P.~G.,  2024, Technical Report
  OSMA/MEO/Report-13, The Activity Profiles and Peak Flux of Radar Meteor
  Showers, \url {https://ntrs.nasa.gov/citations/20240005599}.
NASA Meteoroid Environment Office, \url
  {https://ntrs.nasa.gov/citations/20240005599}

\bibitem[\protect\citeauthoryear{Moorhead, Cooke, Brown  \&
  Campbell-Brown}{Moorhead et~al.}{2025}]{Moorhead2025}
Moorhead A.~V.,  Cooke W.~J.,  Brown P.~G.,   Campbell-Brown M.~D.,  2025,
  \mn@doi [Advances in Space Research]
  {https://doi.org/10.1016/j.asr.2024.08.012}, 75, 1145

\bibitem[\protect\citeauthoryear{Moser \& Cooke}{Moser \&
  Cooke}{2004}]{Moser2004}
Moser D.~E.,  Cooke W.~J.,  2004, \mn@doi [Earth, Moon, and Planets]
  {10.1007/s11038-005-3185-7}, 95, 141

\bibitem[\protect\citeauthoryear{{Moser} \& {Cooke}}{{Moser} \&
  {Cooke}}{2008}]{Moser2008}
{Moser} D.~E.,  {Cooke} W.~J.,  2008, \mn@doi [Earth Moon and Planets]
  {10.1007/s11038-007-9159-1}, \href
  {http://adsabs.harvard.edu/abs/2008EM%26P..102..285M} {102, 285}

\bibitem[\protect\citeauthoryear{{Pittichov{\'a}}, {Woodward}, {Kelley}  \&
  {Reach}}{{Pittichov{\'a}} et~al.}{2008}]{Pittichova2008}
{Pittichov{\'a}} J.,  {Woodward} C.~E.,  {Kelley} M.~S.,   {Reach} W.~T.,
  2008, \mn@doi [Astronomical Journal] {10.1088/0004-6256/136/3/1127}, \href
  {http://cdsads.u-strasbg.fr/abs/2008AJ....136.1127P} {136, 1127}

\bibitem[\protect\citeauthoryear{{Rendtel}, {Vida}  \& {Sugimoto}}{{Rendtel}
  et~al.}{2024}]{Rendtel2024}
{Rendtel} J.,  {Vida} D.,   {Sugimoto} H.,  2024, WGN, Journal of the
  International Meteor Organization, \href
  {https://ui.adsabs.harvard.edu/abs/2024JIMO...52..156R} {52, 156}

\bibitem[\protect\citeauthoryear{{Rendtel}, {Sato}, {Maslov}  \&
  {Vaubaillon}}{{Rendtel} et~al.}{2025}]{Rendtel2025}
{Rendtel} J.,  {Sato} M.,  {Maslov} M.,   {Vaubaillon} J.,  2025, WGN, Journal
  of the International Meteor Organization, 53, 105

\bibitem[\protect\citeauthoryear{Sekanina}{Sekanina}{1985}]{Sekanina1985}
Sekanina Z.,  1985, The Astronomical Journal, 90

\bibitem[\protect\citeauthoryear{Singh, Huebner, Costa, Landaberry  \&
  Pacheco}{Singh et~al.}{1997}]{Singh1997}
Singh P.~D.,  Huebner W.~F.,  Costa R. D.~D.,  Landaberry S. J.~C.,   Pacheco
  J. A. d.~F.,  1997, Planet. Space Sci., 45

\bibitem[\protect\citeauthoryear{{Vaubaillon}, {Watanabe}, {Sato}, {Horii}  \&
  {Koten}}{{Vaubaillon} et~al.}{2011}]{Vaubaillon2011}
{Vaubaillon} J.,  {Watanabe} J.,  {Sato} M.,  {Horii} S.,   {Koten} P.,  2011,
  WGN, Journal of the International Meteor Organization, \href
  {http://cdsads.u-strasbg.fr/abs/2011JIMO...39...59V} {39, 59}

\bibitem[\protect\citeauthoryear{{Vida}, {Campbell-Brown}, {Brown}, {Egal}  \&
  {Mazur}}{{Vida} et~al.}{2020}]{Vida2020}
{Vida} D.,  {Campbell-Brown} M.,  {Brown} P.~G.,  {Egal} A.,   {Mazur} M.~J.,
  2020, \mn@doi [\aap] {10.1051/0004-6361/201937296}, \href
  {https://ui.adsabs.harvard.edu/abs/2020A&A...635A.153V} {635, A153}

\bibitem[\protect\citeauthoryear{Vida, Egal, Brown, Campbell-Brown, Cooke  \&
  Moser}{Vida et~al.}{2024}]{CBET2024}
Vida D.,  Egal A.,  Brown P.~G.,  Campbell-Brown M.~D.,  Cooke W.,   Moser D.,
  2024, 2024 Draconid Meteor Outburst, Central Bureau Electronic Telegrams, No.
  5456, \url {http://www.cbat.eps.harvard.edu/}

\bibitem[\protect\citeauthoryear{{Watson}}{{Watson}}{1934}]{Watson1934}
{Watson} Jr. F.,  1934, Harvard College Observatory Bulletin, \href
  {http://adsabs.harvard.edu/abs/1934BHarO.895....9W} {895, 9}

\bibitem[\protect\citeauthoryear{{Ye}, {Wiegert}, {Brown}, {Campbell-Brown}  \&
  {Weryk}}{{Ye} et~al.}{2014}]{Ye2014}
{Ye} Q.,  {Wiegert} P.~A.,  {Brown} P.~G.,  {Campbell-Brown} M.~D.,   {Weryk}
  R.~J.,  2014, \mn@doi [Monthly Notices of the Royal Astronomical Society]
  {10.1093/mnras/stt2178}, \href
  {http://adsabs.harvard.edu/abs/2014MNRAS.437.3812Y} {437, 3812}

\bibitem[\protect\citeauthoryear{Yeomans \& Chodas}{Yeomans \&
  Chodas}{1989}]{Yeomans1989}
Yeomans D.~K.,  Chodas P.~W.,  1989, The Astronomical Journal, 98, 1083

\makeatother
\end{thebibliography}

\end{document}